\documentclass{IEEEtran}
\usepackage{cite}
\usepackage{amsmath,amssymb,amsfonts}
\usepackage{graphicx}
\usepackage{textcomp,nicefrac}
\usepackage{url}

\usepackage{xcolor}

\def\BibTeX{{\rm B\kern-.05em{\sc i\kern-.025em b}\kern-.08em
T\kern-.1667em\lower.7ex\hbox{E}\kern-.125emX}}
\begin{document}

\thispagestyle{empty}
\twocolumn[
\begin{@twocolumnfalse}

This version of the article has been accepted for publication in the IEEE Transactions on Nuclear Science.\\

\copyright 2025 IEEE.  Personal use of this material is permitted. Permission from IEEE must be obtained for all other uses, in any current or future media, including reprinting/republishing this material for advertising or promotional purposes, creating new collective works, for resale or redistribution to servers or lists, or reuse of any copyrighted component of this work in other works.\\

DOI: 10.1109/TNS.2025.3638275\\
URL: \url{https://doi.org/10.1109/TNS.2025.3638275}\\

Cite: M. Adamič et al., "Readout noise of digital frequency multiplexed TES detectors for CUPID," in \textit{IEEE Transactions on Nuclear Science}, doi: 10.1109/TNS.2025.3638275.\\ 

\end{@twocolumnfalse}
]
\clearpage
\setcounter{page}{1}

\bstctlcite{IEEEexample:BSTcontrol}

\title{Readout noise of digital frequency multiplexed TES detectors for CUPID}

\author{
Michel Adamič, \IEEEmembership{Member, IEEE}, Joseph Camilleri, Chiara Capelli, Matt Dobbs, Tucker Elleflot, Yury G. Kolomensky, Daniel Mayer, Joshua Montgomery, Valentine Novosad, \IEEEmembership{Senior Member, IEEE}, Vivek Singh, Graeme Smecher, \IEEEmembership{Member, IEEE}, Aritoki Suzuki, and Bradford Welliver

\thanks{Accepted manuscript for publication in IEEE TNS.
This work was supported by the US Department of Energy (DOE), Office of Science under Contract No. DE-AC02-05CH11231 and DE-AC02-06CH11357, the DOE Office of Science, Office of Nuclear Physics under Contract No. DE-FG02-00ER41138, and the DOE Office of Science, Office of High Energy Physics. J. Camilleri is supported by the DOE Office of Science Graduate Student Research (SCGSR) program, administered by the Oak Ridge Institute for Science and Education for the DOE under contract number DE-SC0014664.
The McGill team acknowledges funding from the Natural Sciences and Engineering Research Council of Canada and the Canadian Institute for Advanced Research, and the Canada Research Chairs program. Corresponding authors:
}

\thanks{M. Adamič is with the Department of Physics, McGill University, Montreal, QC H3A 2T8 Canada and the Nuclear Science Division, Lawrence Berkeley National Laboratory, Berkeley, CA 94720 USA (e-mail: michel.adamic@mail.mcgill.ca).}
\thanks{J. Camilleri is with the Center for Neutrino Physics, Virginia Polytechnic Institute and State University, Blacksburg, VA 24061 USA and the Nuclear Science Division, Lawrence Berkeley National Laboratory, Berkeley, CA 94720 USA (e-mail: jcamilleri@vt.edu).}
}

\maketitle

\begin{abstract}
Superconducting transition-edge sensor (TES) detectors have been the standard in Cosmic Microwave Background experiments for almost two decades and are now being adapted for use in nuclear physics, such as neutrinoless double beta decay searches. In this paper we focus on a new high-bandwidth frequency multiplexed TES readout system developed for CUPID, a neutrinoless double beta decay experiment that will replace CUORE. In order to achieve the high energy resolution requirements for CUPID, the readout noise of the system must be kept to a minimum. Low TES operating resistance and long wiring between the readout SQUID and the warm electronics are needed for CUPID, prompting a careful consideration of the design parameters of this application of frequency multiplexing. In this work, we characterize the readout noise of the newly designed frequency multiplexed TES readout system for CUPID and construct a noise model to understand it. We find that current sharing between the SQUID coil impedance and other branches of the circuit, as well as the long output wiring, worsen the readout noise of the system. To meet noise requirements, a SQUID with a low input inductance, high transimpedance and/or low dynamic impedance is needed, and the wiring capacitance should be kept as small as possible. Alternatively, the option of adding a cryogenic low-noise amplifier at the output of the SQUID should be explored.
\end{abstract}


\section{Introduction}
\label{sec:introduction}
\IEEEPARstart{S}{uperconducting} transition-edge sensors (TES) have been utilized extensively for two decades in very precise measurements of the Cosmic Microwave Background (CMB). Past and current experiments include SPTpol \cite{SPTpol}, SPT-3G \cite{Benson_SPT,Sobrin_SPT} and the BICEP and Keck array \cite{Bicep_Keck_TES} at the South Pole in Antarctica, as well as ACT \cite{ACT_telescope}, POLARBEAR and the Simons Array \cite{Polarbear2_Simons_Array} in the Atacama desert in Chile. Due to a large number of TES bolometers in these receivers (order 1,000 -- 10,000), multiplexed readout schemes are necessary to reduce the cryogenic complexity and heat load in the cold stages.
One widespread and mature multiplexing technique is frequency division multiplexing (fMUX/FDM) \cite{Dobbs_fmux}, which is also the baseline technology for the future LiteBIRD \cite{LiteBIRD} space mission. Other TES readout schemes include time division multiplexing (TDM), which is the baseline technology for CMB-S4 \cite{CMB-S4}, and microwave SQUID multiplexing ($\mu$MUX) deployed at the newly commissioned Simons Observatory \cite{SO_uMUX}.

Due to their large responsivity and low intrinsic noise, TES have also found application as microcalorimeters in dark matter searches such as CDMS \cite{CDMS_readout}, CRESST \cite{CRESST_TES} and the planned TESSERACT \cite{Tesseract_TES} experiment. In contrast to arrays deployed to study the CMB, the number of channels has been lower and multiplexing schemes have not been employed on a massive scale for these projects. However, the number of readout channels will likely grow with the next generation of rare-event experiments on the horizon. Neutrino physics experiments, specifically neutrinoless double beta decay searches, can fall under a similar class of constraints as their dark matter counterparts and therefore TES have a high potential for impact.

In this paper we focus on the development status and readout noise characterization of a high-bandwidth frequency multiplexed TES readout system for CUPID \cite{CUPID:2025avs}, a next-generation neutrinoless double beta decay experiment. CUPID will be deployed in the cryostat of CUORE \cite{Adams_CUORE}, which is a cryogenic (T$\sim$10 mK) calorimetric array of 988 $\text{Te}\text{O}_2$ crystals (active isotope $^{130}\text{Te}$), operated deep underground at LNGS in Italy. CUPID aims to reduce the dominant $\alpha$-induced background in CUORE with 1596 enriched scintillating $\text{Li}_2\text{MoO}_4$ crystals with $^{100}$Mo as the active isotope. $^{100}$Mo was selected primarily because it avoids low-energy background radiation with a high $Q_{\beta\beta}=3034$ keV, has demonstrated $>95\%$ enrichment, and has the capability of discriminating $\alpha/\beta$ radiation via a scintillating/thermal dual readout. To capitalize on these benefits, the ability for CUPID's light detectors to resolve timing pileup of coincident $2\nu\beta\beta$ decays of $^{100}$Mo is critical. This is relevant for CUPID, but not CUORE, because of much faster $2\nu\beta\beta$ decay rate of $^{100}$Mo compared to $^{130}\text{Te}$ \cite{cupidmo}. This ultimately places an upper limit of $\sim 100$ $\mu$s on the risetime of the light detectors for CUPID. Consequently, fast transition-edge sensors are explored as light detectors to replace germanium neutron-transmutation doped (Ge NTD) thermistors used in CUORE. A large number of light detectors are necessary for CUPID ($\sim2,500$) and future CUPID-1T ($\sim10,000$) \cite{CUPID_1T} experiments. This number of channels to instrument, combined with cryogenic and radiopurity constraints, demands multiplexed readout.

\begin{figure*}[t]
\centerline{\includegraphics[width=7.16in]{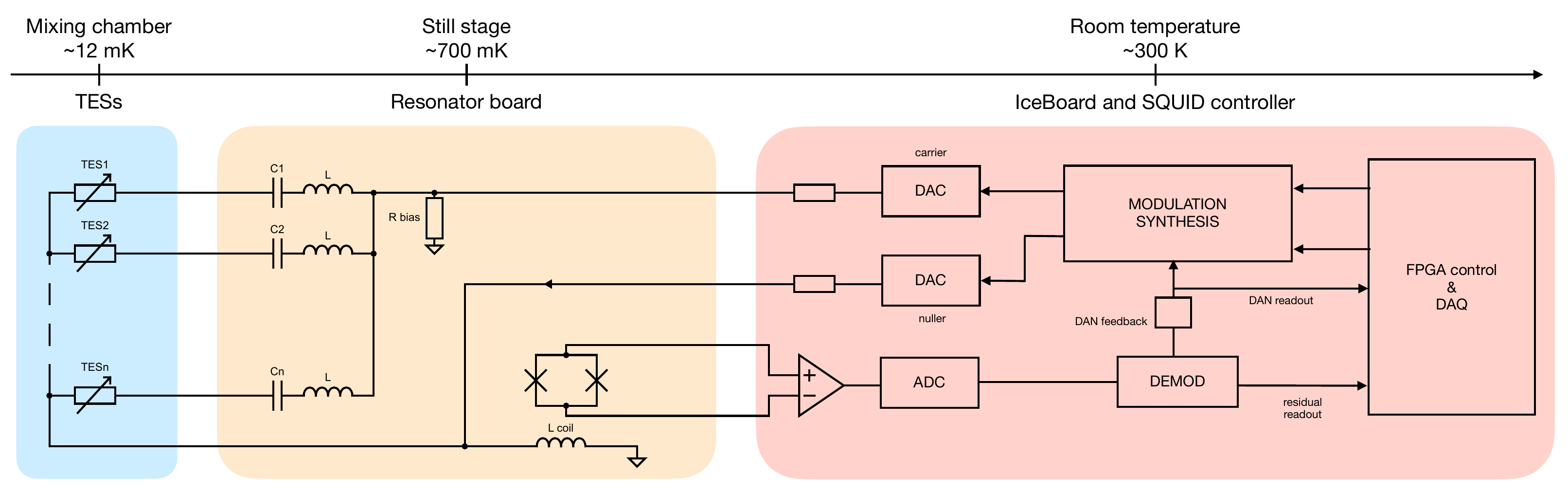}}
\caption{Simplified schematic of the CUPID fMUX TES readout. The bi-layer IrPt TES detectors are on the mixing chamber plate (blue), the resonator bank and readout SQUID on the still plate (yellow), while the ICEboard, its accessories and the analog/digital converters are at room temperature outside the cryostat (red).}
\label{fig:readout_circuit}
\end{figure*}

IrPt bi-layer thin film TES  ($\text{T}_{\text{c}}\sim$ 35 mK) detectors \cite{Singh_TES} were fabricated at Argonne National Lab and characterized at UC Berkeley. To utilize the bandwidth of these detectors, we designed a new high-bandwidth frequency multiplexed TES readout system. The system is based on the SPT-3G readout \cite{Bender_2014} and uses the same Kintex-7 FPGA ICEboard warm readout electronics \cite{Bandura_ICE} from McGill. Since CUPID detectors require a larger bandwidth per channel (several kHz) compared to SPT (76 Hz), both the superconducting multiplexing board and the room-temperature FPGA firmware were redesigned. The multiplexing factor was reduced to 10x (from 128x on SPT-3G \cite{Bender_2014}), while the output data rate was increased dramatically, from 153 samples per second on SPT-3G to 156 ksps for CUPID, three orders of magnitude higher. Fig. \ref{fig:readout_circuit} shows the schematic of the readout system.

Each TES in series with an LC resonator forms a leg of the parallel resonant filter bank, with resonances spaced between 1--5 MHz. The readout firmware synthesizes a set of on-resonance carrier tones (i.e. a carrier comb) to voltage bias each TES in the array. When the TES sensors are heated due to the science signal, the change in TES resistance causes a modulation of the current through each leg. The TES currents are summed and read out with an inductively coupled DC-SQUID transimpedance amplifier. The signal is digitized with a 20 MSPS analog-to-digital converter and sent to the FPGA where it is demodulated to individual channels. In practice, the system operates in the digital active nulling (DAN) \cite{Tijmen_DAN} regime, where the firmware actively cancels any current going through the SQUID input coil through a separate ``nuller'' line, also shown in Fig. \ref{fig:readout_circuit}. This feedback ensures the SQUID stays in the linear operating regime, and as a result, the TES readout currents are flowing through the nuller line. In other words, the nuller line contains the science signal when DAN is active. The total signal bandwidth is limited by the speed of the DAN loop, currently around 3 kHz, just enough to capture $\sim$120 $\mu$s risetimes of the detectors.

The implementation details of the above system, together with the new cold electronics, readout firmware design and first performance measurements of CUPID TES detectors will be presented elsewhere. Here, we focus on the readout noise performance of the system. The structure of the paper is as follows: in Sec. \ref{sec:model}, we present the readout noise model we developed to understand the noise performance of the system. In Sec. \ref{sec:results}, we compare the model to measured data and explore the predictions of the model to varying system parameters. Lastly, we conclude in Sec. \ref{sec:conclusion} with suggestions on how to reduce the readout noise of frequency multiplexed TES detector systems for CUPID-like experiments in the future.

\section{Readout noise model}
\label{sec:model}
To study the readout noise of the system, the detectors are biased above the superconducting transition into the normal region to remove any sensitivity to photons or phonons and DAN is enabled. This results in a readout current on the nuller line, with baseband white noise spectra up to the DAN bandwidth for each channel. The noise for each channel is measured as noise equivalent current, or NEI, in  $\mathrm{pA}/\sqrt{\mathrm{Hz}}$ and depends on the bias frequency. We built a noise model of our system which describes this dependence, and present it below. Some parts of this noise model are derived from the SPT-3G model \cite{SPT_noise_xtalk}, since the same ICEboard warm electronics platform is used in both cases.

SQUID-based frequency multiplexed readout with DAN feedback has two particular noise-amplifying effects, which are crucial for the noise performance of the system. The first is nuller \textit{current sharing} and the second is SQUID \textit{output filtering}, each is described below.

\subsection{Current sharing}
\begin{figure}[h]
\centerline{\includegraphics[width=3.5in]{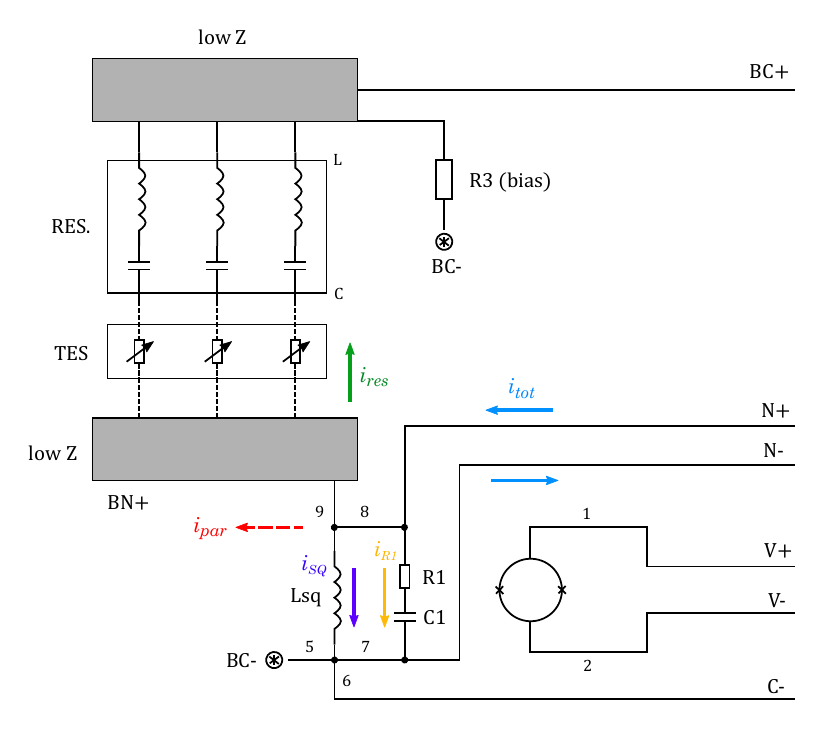}}
\caption{Circuit schematic of the cold electronics board with the SQUID and LC resonators, showing current sharing paths for the nuller (blue arrows). Ideally, all the nuller current would flow through the SQUID input coil (purple).}
\label{fig:sharing_circuit}
\end{figure}

This effect was very problematic for SPT-3G when it deployed and was first reported in \cite{SPT_currentsharing_first}, then later expanded on in \cite{SPT_noise_xtalk}. Looking at Fig. \ref{fig:readout_circuit}, consider a noise source on the demod line of the warm readout, for example at the first stage amplifier after the SQUID. This noise is picked up by the DAN controller in the FPGA, which then injects an inverted ``antinoise'' current on the nuller line, flowing through the SQUID input coil, in an attempt to cancel the demod noise source. However, the injected nuller current doesn't only flow through the SQUID input coil; it also flows around through the TES resonator network, completing the circuit through the 20 m$\Omega$ bias resistor. This means the nuller ``antinoise" current has to be that much higher to cancel the demod noise source, effectively boosting the readout noise by the \textit{current sharing factor, csf}. Fig. \ref{fig:sharing_circuit} shows the multiplexing board schematic with a complete path of the nuller current through the cold electronics. Note that the signal lines are differential, not single-ended as depicted in the simplified schematic in Fig. \ref{fig:readout_circuit}.

The nuller current $i_{tot}$ is split into four branches: the desired SQUID input coil current $i_{SQ}$, current through the parallel snubber resistor $i_{R1}$, current through the TES resonator network $i_{res}$ and finally the parasitic current $i_{par}$ leaking through capacitive coupling to ground. We can then define the current sharing factor $csf$ or $1/\chi_{cs}$ \cite{SPT_noise_xtalk,Joshua_PhD} as
\begin{equation}
\label{eq:csf}
csf \equiv \frac{1}{\chi_{cs}} \equiv \frac{i_{tot}}{i_{SQ}} = \left| 1+j\omega L_{SQ}\left(\frac{1}{R_1}+Y_{res}+Y_{par}\right)\right| ,
\end{equation}
where $j\omega L_{SQ}$ is the SQUID input impedance and $Y_{res}$ and $Y_{par}$ are resonator and parasitic admittances ($1/Z$). Ideally, if the SQUID input impedance was very low compared to other branches of the circuit, most of the current would flow through the SQUID coil, but that is not always the case.

\begin{figure}[h]
\centerline{\includegraphics[width=3.5in]{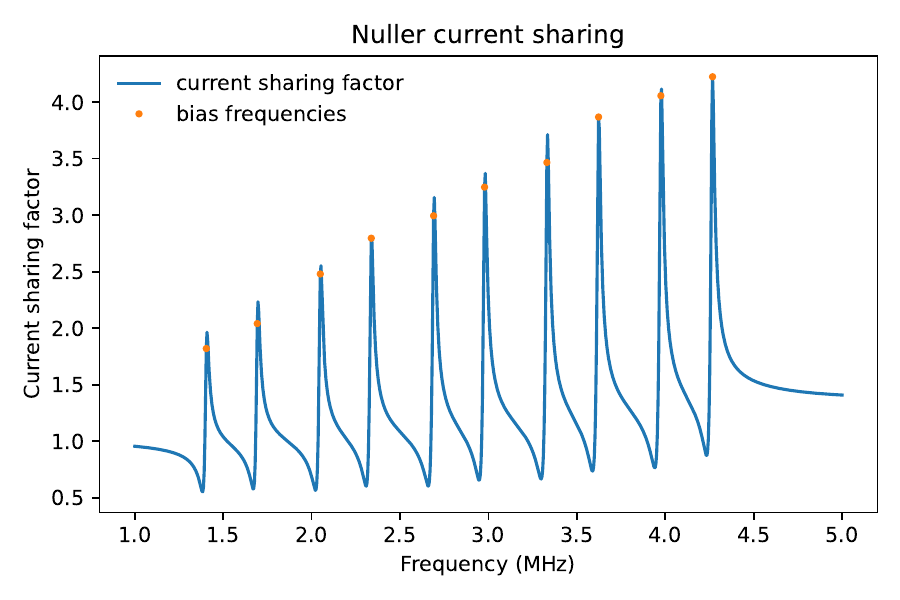}}
\caption{Modeled current sharing factor for our system parameters. Current sharing increases with frequency due to rising SQUID coil impedance and spikes on-resonance due to low TES operating resistance.}
\label{fig:current_sharing}
\end{figure}

Our SQUID is a NIST SA13 DC SQUID array \cite{sa13} with an estimated input inductance of 80 nH, resulting in an impedance of 0.5 -- 2.5 $\Omega$ in the 1 -- 5 MHz range of bias frequencies. $R_1$ is a 5 $\Omega$ SQUID snubber resistor to remove the irregularities from the SQUID's response curve at low temperature \cite{Tucker_noise}. However, its current sharing effect is low compared to the resonator network, because the TES are operated close to 0.5\,$\Omega$ of resistance in the transition. Therefore, (\ref{eq:csf}) predicts the current sharing factor to spike at resonance frequencies (i.e. where the TES are ``visible") and the situation to worsen as the SQUID coil impedance rises. This matches what we observe if we model the current sharing factor for our system -- see Fig. \ref{fig:current_sharing}.

The last term in (\ref{eq:csf}), which corresponds to leakage through parasitic capacitances to ground, was a significant problem for SPT-3G \cite{SPT_noise_xtalk}. However, the SPT team found and proposed a solution: isolating the cold components from cryostat ground with a resistor on the Iceboard's SQUID controller board. We increased our grounding resistors (label $R_{48}$ in the Appendix) to 100 $\Omega$, which is much higher than the SQUID input impedance; therefore, the parasitic current for our setup is negligible.

\subsection{SQUID output filter}
The second noise boosting effect we need to analyze is the SQUID output filter due to the long cabling between the output of the SQUID and the first stage amplifier on the warm electronics. The SQUID is installed on the still stage inside of the cryostat (see Fig. \ref{fig:readout_circuit}), so some length of twisted pair cable is needed to transfer the output signal to the warm electronics. This wire harness has capacitance between the conductors $C_{wh}$, while the output of the SQUID has a certain dynamic impedance, defined as
\begin{equation}
\label{eq:Z_dyn}
Z_{dyn} = \frac{\partial V_{out}}{\partial I_{bias}} ,
\end{equation}
where $V_{out}$ and $I_{bias}$ are the voltage and current on the output side of the SQUID. This impedance, together with the cable capacitance, forms a RC low-pass filter that attenuates the signal coming from the SQUID. If we now have a noise source on the demod line of the readout (i.e. after the filter), it will translate to a higher equivalent noise source on the input of the SQUID due to signal attenuation, effectively boosting the readout noise. We can write the magnitude response of the filter transfer function \cite{SPT_noise_xtalk,Joshua_PhD} as
\begin{equation}
\label{eq:filter}
\chi_{out} \equiv \frac{V_{Amp}}{V_{out}} \approx \left| \frac{1}{1+j\omega Z_{dyn}C_{wh}}\right|,
\end{equation}
which is a standard first order low pass filter between the SQUID output voltage $V_{out}$ (input of the filter) and the voltage at the first stage amplifier $V_{Amp}$ (output of the filter). The expression is technically an approximation because a long cable is not a discrete capacitor, but a distributed line of capacitance, inductance and possibly resistance on the non-superconducting end of the wire. We performed simulations in LTSpice for our cable parameters comparing the simple capacitor model with a more complex LCR line, and the difference in magnitude and phase transfer functions was barely noticeable. Therefore, we conclude the RC filter approximation in (\ref{eq:filter}) is adequate for the purposes of our noise model.

In our experimental setup we have 0.8 meters of superconducting GVL Cryoengineering GVLZ289 twisted pair Nb-Ti coaxial cable between the SQUID and the 4K stage with a measured capacitance of 35 pF. Then, an additional 1.0 meters of Cryoloom twisted pair copper cable leading from the 4K stage to the warm electronics, with a measured capacitance of 85 pF. In total the cabling presents a 120 pF capacitive load to the SQUID, which we tuned to a dynamic impedance of $Z_{dyn} = 820$ $\Omega$. This results in an output filtering $\chi_{out}$ response shown in Fig. \ref{fig:output_filter} below.

\begin{figure}[h]
\centerline{\includegraphics[width=3.5in]{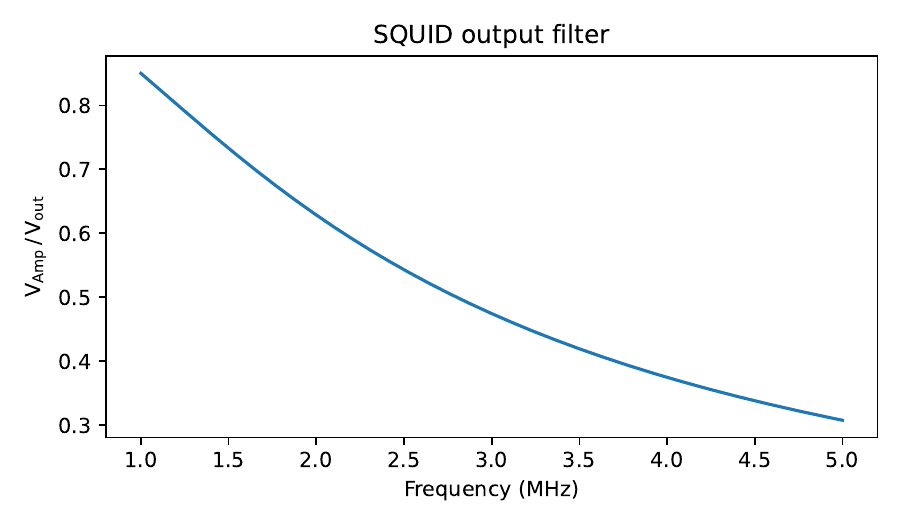}}
\caption{Modeled SQUID output filter magnitude response for a dynamic impedance of 820 $\Omega$ and wire harness capacitance of 120 pF.}
\label{fig:output_filter}
\end{figure}

The attenuation is quite severe, highlighting that this effect is something that needs to be addressed and minimized when designing the full readout system for CUPID. Looking at the readout cabling solution presently used for CUORE \cite{cuore_electronics}, the distance from the SQUID to the warm electronics would be around 4 meters with a capacitance of 400 pF, which is unacceptable for TES fMUX readout. A potential borderline solution would be to either use low capacitance wiring like our NbTi at 40 pF/m, resulting in $C_{wh}\approx160$ pF, or to shorten the wiring. The first stage amplifiers could potentially be placed right above the cryostat, roughly 1.5 meters from the still stage \cite{cuore_cryostat}, resulting in $C_{wh}\approx60-150$ pF depending on the type of cable used. In case none of that is feasible, additional cryogenic amplification stages will be needed inside the cryostat to avoid the output filtering effect.

\subsection{Noise sources}
Now that we understand the noise amplifying terms in the system, we can add the actual noise sources in our model. The modeled noise sources have a white spectrum, corresponding to various resistors and amplifiers in our electronics. Following the list from the SPT-3G readout \cite{Joshua_PhD}, the main noise sources are the NIST SA13 SQUID and various components from the warm electronics, namely the first stage amplifier (Linear Technology LT6200-5), the SQUID bias resistor, nuller stiffening resistors, resistors on the demod signal path, the ADC \& DACs etc. The full list can be found in Table \ref{tab1} in the Appendix, alongside the parameters and equations used for the noise model. It is worth emphasizing that the terms in the demod path of the signal chain have the current sharing $(\chi_{cs})$ and output filtering $(\chi_{out})$ factors in their respective equations, boosting those noise sources. In addition, every noise source is also multiplied by the modulation penalty $\chi_{mod} = \sqrt{2}$ due to the sinusoidal modulation of the signals \cite{Joshua_PhD}. Lastly, all modeled noise sources are added in quadrature to get the total readout noise of the system.

\section{Results}
\label{sec:results}
\begin{figure}[h]
\centerline{\includegraphics[width=3.5in]{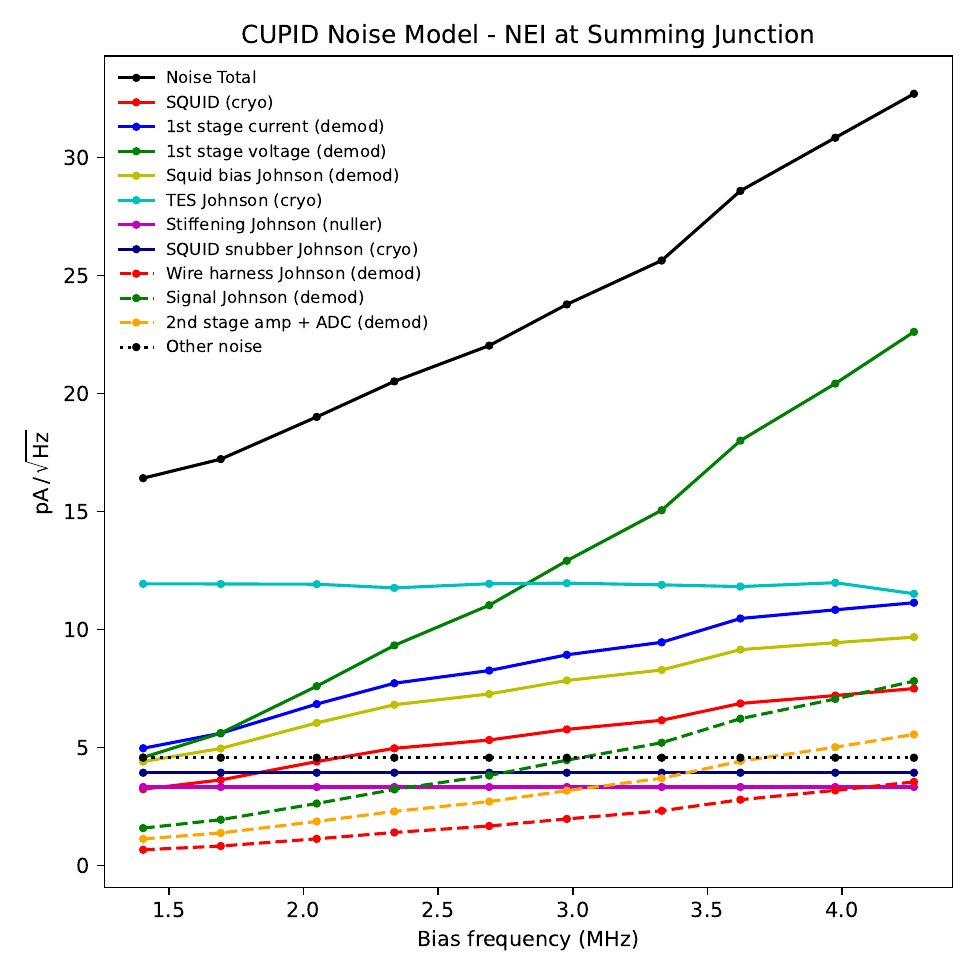}}
\caption{On-resonance readout noise model prediction for our system parameters, showing contributions from different noise sources and the total noise in black, including the TES Johnson noise.}
\label{fig:noise_model}
\end{figure}

Fig. \ref{fig:noise_model} shows the on-resonance noise model prediction for the prototype CUPID frequency multiplexed readout system, operated above the superconducting transition with a normal TES resistance of $0.5$ $\Omega$. The ten points correspond to the resonant frequencies of our detectors. The total noise in the operating frequency range is between 15--35 $\mathrm{pA}/\sqrt{\mathrm{Hz}}$ (referred to the SQUID input), rising monotonically with bias frequency. The largest contributor is the first stage amplifier on the Iceboard SQUID controller board (warm electronics), followed by the SQUID bias resistor and the SQUID itself. Interestingly, as suspected, the noise sources that rise with frequency and at the same time contribute the most to the total noise are all on the demod line of the readout signal path (i.e. the SQUID itself and everything following it), because they are boosted by the frequency-dependent current sharing and the output filtering effects. Other noise sources do not have a frequency dependence, as expected for white noise sources.

Note that we intentionally ignored the Johnson noise contribution from the TES resistance at around 12 $\mathrm{pA}/\sqrt{\mathrm{Hz}}$ (also shown in Fig. \ref{fig:noise_model}), since, even though it is included in the noise model, does not technically count towards the readout noise of the system. That is because when the TES are operating in a real experiment, they are kept under electro-thermal feedback (ETF), which  cancels TES Johnson noise, replacing it with detector phonon and photon noise contributions. These depend on the operation of the detectors themselves and are not part of this paper's discussion.

\subsection{Model prediction vs experimental data}
To validate the accuracy of the noise model, we first compared it to the reported noise for SPT-3G \cite{Joshua_PhD,SPT_noise_xtalk}, since our readout system is in essence very similar. For the parameters of the SPT-3G experiment, our model yields virtually the same noise prediction, including all the subcomponents, of the noise reported by SPT-3G \cite{Joshua_PhD,SPT_noise_xtalk}, alongside with very similar plots for current sharing and SQUID output filtering. This serves as a good self-consistency check for our noise model, even though we model some critical components (such as current sharing) differently from the SPT-3G collaboration.

\begin{figure}[h]
\centerline{\includegraphics[width=3.5in]{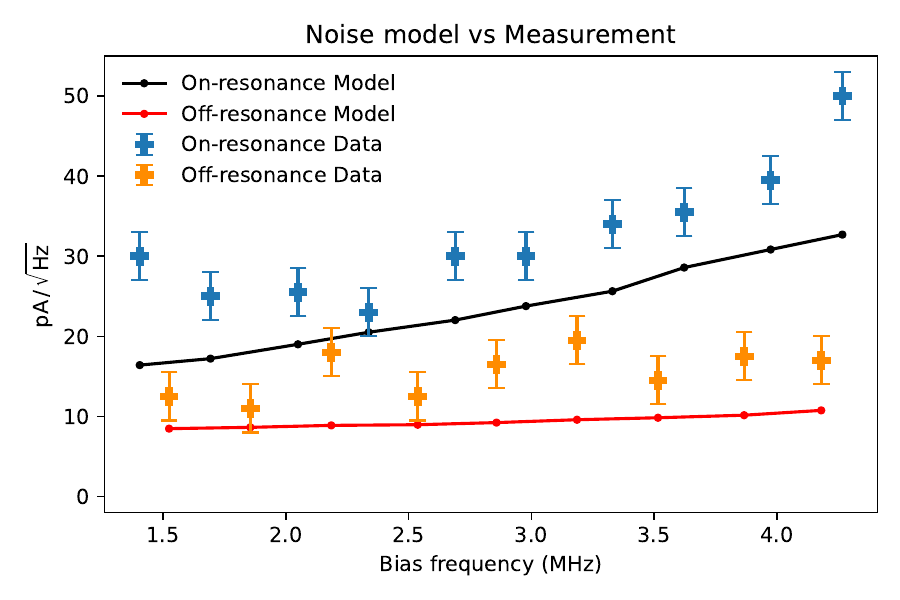}}
\caption{Measured on- and off-resonance noise vs model prediction for $0.5$ $\Omega$ ``dummy TES'' (i.e. resistors) and SQUID $Z_{dyn}=820$ $\Omega$.}
\label{fig:single_data}
\end{figure}

Next, we performed noise measurements with our experimental setup at UC Berkeley. For the noise characterization of the system, we replaced each TES with $0.5$ $\Omega$ SMD resistors, since we are not interested in TES noise and biasing regular resistors is much easier. This is also what was assumed in the model in Fig. \ref{fig:noise_model}. The resistors were soldered directly to the SQUID resonator board and cooled down to 700 mK. We are using thin metal film resistors with a very small temperature coefficient that have been previously measured with a 4-point setup to have virtually the same resistance at 1 K as at room temperature. We biased our NIST SA13 SQUID array on the negative slope of the $V(\phi)$ curve with a measured transimpedance and dynamic impedance of 820 $\Omega$ ($Z_{trans}=820$ $\Omega$, $Z_{dyn}=820$ $\Omega$). Lastly, the carrier tones on all 10 resonance frequencies were turned on with a normalized amplitude of 0.003 of the DAC dynamic range and the DAN loop was enabled with a bandwidth of 3 kHz. Noise values were obtained by taking the median value of the flat part of the noise spectrum for each channel, i.e. between 0 -- 3 kHz. We repeated the measurement multiple times and took the average to get each datapoint. The results of on-resonance and off-resonance noise measurements, compared to the model predictions, are shown in Fig. \ref{fig:single_data}.

First of all, the noise off-resonance is lower than on-resonance, since the current sharing factor is much lower between resonances (see Fig. \ref{fig:current_sharing}) and the TES resistor Johnson noise is not present. Secondly, we notice that noise datapoints deviate from the monotonic trend, and they are consistently higher from the model prediction by about 5 $\mathrm{pA}/\sqrt{\mathrm{Hz}}$. We are not sure why that is, but suspect either external noise sources or some missed component in the overall gain of the system that is multiplying the noise. Initially we had much higher noise levels, but managed to bring them down progressively by optimizing the grounding scheme of our system, which suggests that we are still picking up some unwanted noise from the environment. The lowest noise levels we achieved (presented here) were obtained by using the SPT/LiteBIRD grounding scheme, i.e. by grounding the cryostat directly to the Iceboard's power supply ground, a common grounding point for the whole system. We hope that with additional shielding and further ground optimizations we will be able to bring the noise levels closer to the model prediction. Since the trends for both on- and off-resonance noise agree with data and are consistent with SPT-3G measurements, we have pretty strong confidence in the model.

\subsection{Measured noise vs $Z_{dyn}$}
\begin{figure}[h]
\centerline{\includegraphics[width=3.5in]{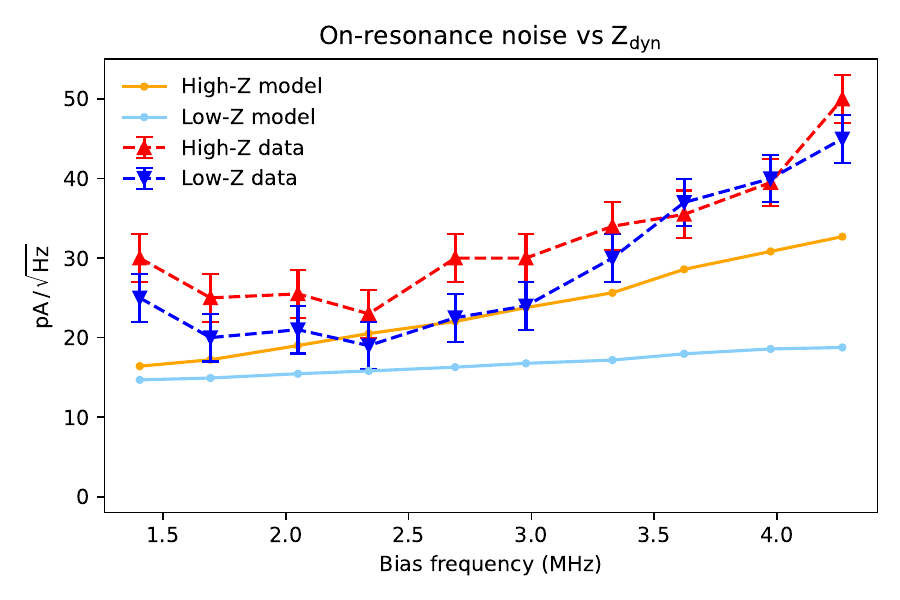}}
\caption{Measured on-resonance noise vs model prediction for low ($Z_{dyn}=260$ $\Omega$) and high ($Z_{dyn}=820$ $\Omega$) SQUID operation.}
\label{fig:Zdyn_data}
\end{figure}

Next, we biased the SQUID on the positive slope of the $V(\phi)$ curve, where the dynamic impedance is much lower ($Z_{dyn}=260$ $\Omega$), while maintaining high transimpedance ($Z_{trans}=950$ $\Omega$). This stable bias point, successfully reported in \cite{Tucker_noise}, is the preferred operation of the NIST SA13 SQUID at this temperature and is achievable because of the 5 $\Omega$ snubber resistor that removes the irregularities on the positive slope of the SQUID's $V(\phi)$ curve. We repeated the noise measurement with this lower $Z_{dyn}$ setting, expecting to see lower noise due to reduced output filtering effect. The on-resonance results are shown in Fig. \ref{fig:Zdyn_data}, comparing them to the $820$ $\Omega$ SQUID bias point before.

Even though the noise dropped slightly, the improvement is not as good as predicted, increasing the disagreement between the data and the model. This suggests that we are definitely seeing excess noise in the system. In the following subsection, we present additional experiments we performed to try to identify the excess noise component.

\subsection{Measured noise vs DAN gain, carrier ampltitude and DAC gain, number of tones, frequency placement etc.}
We repeated the noise measurement by operating the Iceboard readout system with different settings, to rule out any missed noise contributions from the warm electronics.

To make sure we are exercising enough dynamic range of the carrier and nuller digital-to-analog converters (DACs), we performed noise measurements at different carrier amplitudes, spanning from 0.001 to 0.03 of normalized amplitude (a range of 30x) per channel. This did not result in any change in white noise levels. Similarly, adding a large-amplitude (0.1 norm.) dither carrier tone away from the resonances had no effect. Next, we operated the DAC amplifiers on the Iceboard mezzanine cards at different gain settings, from 0 to 15, again with no change in noise whatsoever. We first suspected that the DAC amplifier chain on our mezzanines with the ``PB2" filtering configuration might be responsible for the additional noise. However, models suggest a strong dependence in noise on the DAC amplifier gain, which we have not observed. Therefore, even though we are still investigating this scenario, the data strongly disfavors the DAC chain as the source of the excess noise.

Next, we operated the readout at different DAN gains, from 1 kHz to 3 kHz of bandwidth, to rule out the effect of DAN antinulling \cite{Smecher_DAN} on noise. No change was observed, showing that our DAN loop is stable and operating as intended without injecting any additional noise in the SQUID.

Lastly, to rule out any interference between carrier tones or the effect of intermodulation distortion (IMD) products, we performed noise measurements with only one active channel at a time, i.e. without multiplexing. This reduced the number of IMD spikes in the high frequency part of the spectrum, but had no effect on the white noise levels. Similarly, we set the readout tone frequencies to multiples of 78125 Hz (output sampling Nyquist frequency) to remove most of the IMD spikes, again with no significant change in white noise levels. To make sure we have no statistical contamination in the white noise data, we checked the Gaussianity of the spectra used to calculate noise levels. The data have nice Gaussian distributions and the median is doing a good job keeping the extracted noise values at the center of the distributions, despite some spikes at higher noise levels.

From these tests we conclude that the readout system is operating as intended and we have probably not missed any significant noise contributions from the warm electronics. We assume that we have either an anomalous gain that affects the way noise is referred to the input of the SQUID or we are seeing excess noise from the environment, which is inflating our white noise levels at baseband. We are actively investigating this and trying to shield our system better from external RFI.

\subsection{Path for improved noise performance}
\begin{figure}[h]
\centerline{\includegraphics[width=3.5in]{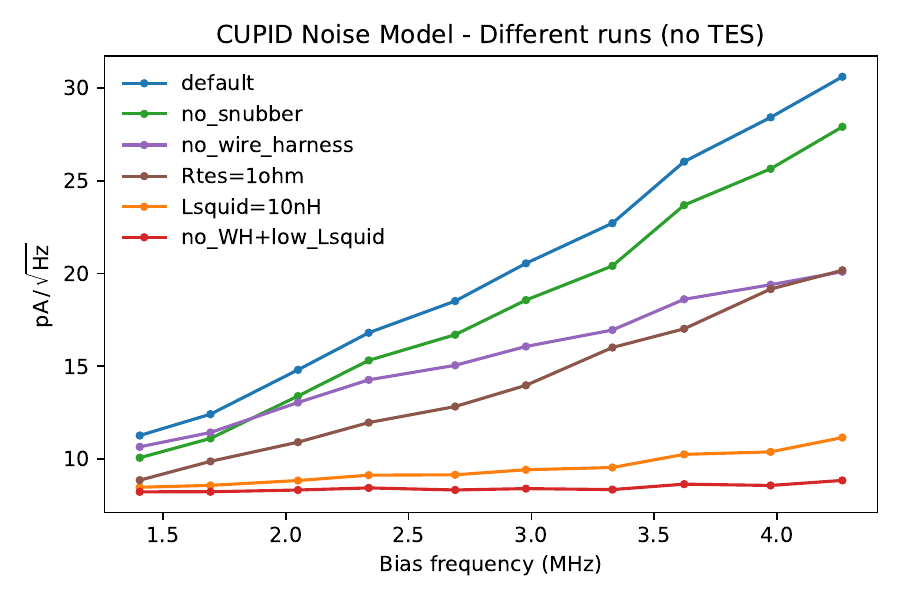}}
\caption{CUPID on-resonance noise model runs for different scenarios at $Z_{dyn}=Z_{trans}=820$ $\Omega$ SQUID operation excluding TES Johnson noise. All changes from the default baseline are applied one at a time, except in the lowest noise scenario, where two changes are applied together.}
\label{fig:noise_comparisons}
\end{figure}

Based on the understanding of the readout system and the cryogenic architecture provided by our model, we can make predictions of how to reduce the noise in future iterations of the hardware. We take the model in Fig. \ref{fig:noise_model} as the baseline (i.e. with $Z_{dyn}=Z_{trans}=820$ $\Omega$), but this time excluding the TES Johnson noise contribution, since we are interested purely in the readout noise. We change parameters of the model to evaluate improvements to readout noise for different realistic changes that can be implemented. Fig. \ref{fig:noise_comparisons} shows the results of this.

First, removing the 5 $\Omega$ SQUID snubber resistor (to reduce current sharing) has a small effect and is not recommended, since the snubber allows us to operate the SQUID in the low dynamic impedance regime, where the noise reduction is much better. The effect of very low $Z_{dyn}$ operation is shown by the purple line, where the SQUID output filtering effect has been removed by reducing the wire harness capacitance to be negligible. The output filtering effect is very important for CUPID, because we are dealing with large wiring capacitances due to long cables between the SQUID and the first stage amplifier in the warm electronics. As shown here, ``disabling'' this filter (as is the case for the short wiring harness used in the multiplexed readout implementation for SPT-3G \cite{SPT_noise_xtalk}) would improve the readout noise significantly.

Even bigger noise reductions are achievable if we reduce the current sharing effect. The first option is to increase the TES operating resistance, which is currently $0.5$ $\Omega$ or even lower. If we increase this to $1$ $\Omega$ (similar to the impedance of the SPT detectors \cite{Sobrin_SPT}), we already see a drastic reduction in noise (brown line). This calls for a potential redesign of the developed TES sensors provided we decide to keep using the NIST SA13 SQUID for frequency multiplexed readout. The current TES design is not set in stone, so developing TESs with $R\sim1\,\Omega$ of operating resistance is a viable option to consider for CUPID.

However, none of this matters too much if we can employ a SQUID with a 10 nH input inductance (shown in orange), assuming the transimpedance remains high (for example, StarCryo has developed SQUIDs with these parameters \cite{boyd2017_SQUID}). This basically eliminates current sharing, dropping the noise to 10 $\mathrm{pA}/\sqrt{\mathrm{Hz}}$ with very slight frequency dependence due to output filtering. If we combine such a SQUID with no wiring capacitance (red), removing both current sharing and output filtering effects, the noise loses frequency dependence and is 8 $\mathrm{pA}/\sqrt{\mathrm{Hz}}$ for all channels.

Very low noise operation of frequency multiplexed TES sensors with the Iceboard hardware has been demonstrated by the LiteBIRD group at Berkeley, where 8--10 $\mathrm{pA}/\sqrt{\mathrm{Hz}}$ noise levels have been achieved with very low frequency dependence \cite{Tucker_noise}. There, all cryogenic components are cooled down to 250 mK, and the SA13 SQUID is operated at low $Z_{dyn}$ with very short output cabling, removing the output filtering effect. The group also uses 1 $\Omega$ TES resistors. 
By implementing the changes outlined here, we expect our CUPID prototype should be able to approach those noise levels.

\section{Conclusion}
\label{sec:conclusion}
We built a high-bandwidth 10x frequency multiplexed TES readout system for CUPID, using the Iceboard readout electronics and experience from Cosmic Microwave Background experiments. Subtracting TES Johnson noise, the system demonstrates a 15--40 $\mathrm{pA}/\sqrt{\mathrm{Hz}}$ readout noise performance in the 1--4.5 MHz frequency range with $0.5$ $\Omega$ TES resistors and the NIST SA13 SQUID at 700 mK. We developed an extensive readout noise model of the system and identified two main noise boosting effects that cause the noise to increase with carrier frequency, namely the nuller current sharing and the SQUID output filtering, which are already known to the CMB instrumentation community. Both effects are particularly problematic for the CUPID readout, which currently uses low resistance TES sensors (increasing the current sharing) and anticipates long wiring on the output of the SQUID due to the size of the CUPID cryostat. We propose increasing the TES resistance or switching to a lower input inductance SQUID, to minimize current sharing. In addition, the SQUID needs to be operated in the low dynamic impedance regime, while careful considerations are needed regarding the capacitance of the output wiring. If the wiring cannot be made shorter or have much lower capacitance in the final experiment, the option of installing a cryogenic amplifier after the SQUID should be explored.

Meanwhile, the measurements at Berkeley are still showing moderate excess noise compared to the model, especially at low $Z_{dyn}$ and higher frequencies. We suspect either an issue with the modeling of the system gains,
external RFI, or the DAC filter chain, although the latter theory is strongly disfavoured by data at different mezzanine gains. Further investigations are underway, both in terms of theory and new noise measurements. 

\appendices

\section*{Appendix}
Table \ref{tab1} lists all expressions and parameters used in our noise model. Below we also write the equations for the modulation noise penalty ($\chi_{mod}$), the current sharing factor ($\chi_{cs}^{-1}$) and the SQUID output filter ($\chi_{out}$). The model also needs the equivalent resistance of the wire harness and the SQUID dynamic impedance ($R_{eq}$) and the total resistance seen between the input terminals of the 1st stage amplifier ($R_{sqcb}$) -- see Ref. \cite{Joshua_PhD}.
\\
\begin{table*}[h]
\centering
\caption{\label{tab1} Expressions and parameters used to model CUPID readout noise}
\renewcommand{\arraystretch}{1.5} 
\setlength{\tabcolsep}{3pt}
\begin{tabular}{|p{1.55in}|p{2.75in}|p{2.5in}|}
\hline
Noise source & Expression for NEI at SQUID summing junction & Model parameters \\
\hline
SQUID & $\frac{\chi_{mod}}{\chi_{cs}}\sqrt{\frac{T_{squid}}{4 K}}\cdot 3 \frac{\mathrm{pA}}{\sqrt{\mathrm{Hz}}}$ & $R_{bias} = 20$ m$\Omega$ \\

1st stage amp voltage & $\frac{\chi_{mod}}{\chi_{cs}\chi_{out}Z_{trans}}\cdot 1.1 \frac{\mathrm{nV}}{\sqrt{\mathrm{Hz}}}$ & $R_{snubber} = 5$ $\Omega$ \\

1st stage amp current & $\frac{\chi_{mod}}{\chi_{cs}\chi_{out}Z_{trans}}R_{sqcb}\cdot 2.2 \frac{\mathrm{pA}}{\sqrt{\mathrm{Hz}}}$ & $T_{squid} = T_{TES} = 700$ mK \\

SQUID bias Johnson (4.22 k$\Omega$) & $\frac{\chi_{mod}}{\chi_{cs}\chi_{out}Z_{trans}} \frac{R_{eq}}{R_{eq}+4.22k\Omega}\cdot 8.36 \frac{\mathrm{nV}}{\sqrt{\mathrm{Hz}}}$ & $R_{TES} = R_n = 0.5$ $\Omega$ \\

TES Johnson$^{\mathrm{1}}$ & $\chi_{mod}\sqrt{\frac{T_{TES}}{0.1 K}}\cdot \frac{2.35}{\sqrt{R_{TES}}} \frac{\mathrm{pA}}{\sqrt{\mathrm{Hz}}}$ & $L_{squid} = 80$ nH \\

Nuller stiffening Johnson (3 k$\Omega$) & $\chi_{mod}\cdot 2.35 \frac{\mathrm{pA}}{\sqrt{\mathrm{Hz}}}$ & $Z_{dyn} = 820$ $\Omega$,  $Z_{trans} = 820$ $\Omega$ (high $Z_{dyn}$ op.) \\

SQUID snubber Johnson & $\chi_{mod}\sqrt{\frac{T_{squid}}{0.1 K}}\sqrt{\frac{5\Omega}{R_{snubber}}}\cdot 1.05 \frac{\mathrm{pA}}{\sqrt{\mathrm{Hz}}}$ & $Z_{dyn} = 260$ $\Omega$,  $Z_{trans} = 950$ $\Omega$ (low $Z_{dyn}$ op.) \\

Demod wire harness Johnson & $\frac{\chi_{mod}}{\chi_{cs}\chi_{out}Z_{trans}} \frac{4.22k\Omega}{R_{eq}+4.22k\Omega}\sqrt{\frac{T_{wh}}{300 K}}\sqrt{\frac{R_{wh}}{100\Omega}}\cdot 1.29 \frac{\mathrm{nV}}{\sqrt{\mathrm{Hz}}}$ & $T_{wh} = 300$ K, $R_{wh} = 1$ $\Omega$ (wire harness) \\

Demod signal path Johnson & $\frac{\chi_{mod}}{\chi_{cs}\chi_{out}Z_{trans}}\cdot 0.38 \frac{\mathrm{nV}}{\sqrt{\mathrm{Hz}}}$ & $C_{wh} = 120$ pF, $L_{wh} = 800$ nH (wire harness) \\

2nd stage amp + ADC & $\frac{\chi_{mod}}{\chi_{cs}\chi_{out}Z_{trans}}\cdot 0.37 \frac{\mathrm{nV}}{\sqrt{\mathrm{Hz}}}$ & $R_{48} = 100$ $\Omega$ (ground resistor) \\

Other$^{\mathrm{2}}$ & $4.2 \frac{\mathrm{pA}}{\sqrt{\mathrm{Hz}}}$ &  \\

\hline
\multicolumn{3}{p{7in}}{$^{\mathrm{1}}$On-resonance only, disabled off-resonance. ``TES" are just SMD resistors. $^{\mathrm{2}}$Other noise sources from \cite{Joshua_PhD} summed together in quadrature.}
\end{tabular}
\end{table*}

\noindent $\chi_{mod} = \sqrt{2}$ \\
$\chi_{cs}^{-1} = \frac{i_{tot}}{i_{SQ}} = \left\lvert 1 + j\omega L_{SQ}\left(\frac{1}{R_{snubber}}+Y_{res}+\frac{1}{R_{48}}\right) \right\rvert$ \\
$\chi_{out} = \left\lvert \frac{1}{1+j\omega Z_{dyn}C_{wh}} \right\rvert $ \\
$R_{eq} = \left\lvert 2R_{wh} + 2j\omega L_{wh} + \left(\frac{1}{Z_{dyn}}+j\omega C_{wh}\right)^{-1} \right\rvert $ \\
$R_{sqcb} = \left(\frac{1}{10\Omega}+\frac{1}{100\Omega}+\frac{1}{150\Omega}\right)^{-1} + \left(\frac{1}{4.22k\Omega}+\frac{1}{R_{eq}}\right)^{-1} $ \\

\section*{Acknowledgment}
The authors would like to thank Tijmen de Haan and Bill Holzapfel for offering their insights and grounding tips from the SPT world, Nicole Farias for kindly sharing her lab space and noise findings from LiteBIRD, Carl Haber for impedance measurements of the cryogenic cabling we use in our cryostat, and Jean-Francois Cliche for help with noise modeling in LTspice.

\bibliographystyle{IEEEtran}
\bibliography{biblio}

\end{document}